\documentclass[lettersize,journal]{IEEEtran}
\usepackage{amsmath,amsfonts}
\usepackage{array}
\usepackage[caption=false,font=normalsize,labelfont={footnotesize},textfont={footnotesize}]{subfig}
\usepackage{textcomp}
\usepackage{stfloats}
\usepackage{url}
\usepackage{verbatim}
\usepackage{graphicx}
\usepackage{cite}
\usepackage[hidelinks]{hyperref}
\usepackage{ctable}
\usepackage{colortbl}
\usepackage{epstopdf}
\usepackage[ruled,lined,linesnumbered,commentsnumbered,longend]{algorithm2e}
\usepackage{comment}
\usepackage{algpseudocode}
\usepackage{bm}
\usepackage{amsthm}
\theoremstyle{definition}

\newtheorem*{definition*}{Definition}

\begin{document}

\title{Late multimodal fusion for image and audio music transcription}

\author{María Alfaro-Contreras, Jose J. Valero-Mas, José M. Iñesta, Jorge Calvo-Zaragoza
\thanks{The authors are with Instituto Universitario de Investigación Informática, University of Alicante, Ap. 99, 03080 Alicante, Spain. Corresponding author: malfaro@dlsi.ua.es}}

\markboth{Journal of \LaTeX\ Class Files,~Vol.~14, No.~8, August~2021}%
{Shell \MakeLowercase{\textit{et al.}}: A Sample Article Using IEEEtran.cls for IEEE Journals}



\maketitle

\begin{abstract}
Music transcription, which deals with the conversion of music sources into a structured digital format, is a key problem for Music Information Retrieval (MIR). When addressing this challenge in computational terms, the MIR community follows two lines of research: music documents, which is the case of Optical Music Recognition (OMR), or audio recordings, which is the case of Automatic Music Transcription (AMT). The different nature of the aforementioned input data has conditioned these fields to develop modality-specific frameworks. However, their recent definition in terms of sequence labeling tasks leads to a common output representation, which enables research on a combined paradigm. In this respect, multimodal image and audio music transcription comprises the challenge of effectively combining the information conveyed by image and audio modalities. In this work, we explore this question at a late-fusion level: we study four combination approaches in order to merge, for the first time, the hypotheses regarding end-to-end OMR and AMT systems in a lattice-based search space. The results obtained for a series of performance scenarios---in which the corresponding single-modality models yield different error rates---showed interesting benefits of these approaches. In addition, two of the four strategies considered significantly improve the corresponding unimodal standard recognition frameworks.
\end{abstract}


\section{Introduction}
\label{sec:introduction}
\IEEEPARstart{A}{}long-standing research problem in Music Information Retrieval (MIR) is that of attaining structured digital representations from music sources, typically known as (music) \textit{transcription}~\cite{serra2013roadmap}. The MIR community follows two main research lines that study how to computationally solve this problem when targeting either music documents---known as Optical Music Recognition (OMR)~\cite{calvo2020understanding}---or acoustic music signals---namely, Automatic Music Transcription (AMT)~\cite{benetos2018automatic}. Despite having a similar purpose, these two fields have historically evolved in separate ways owing to differences in the nature of the data, which has resulted in specific task-oriented recognition frameworks that are most typically based on multi-stage procedures~\cite{de2021multimodal}.

\begin{figure}[!t]
\centering
\subfloat[Optical Music Recognition: music transcription using score images as inputs and a music-representation language as output.]{\includegraphics[width=\columnwidth]{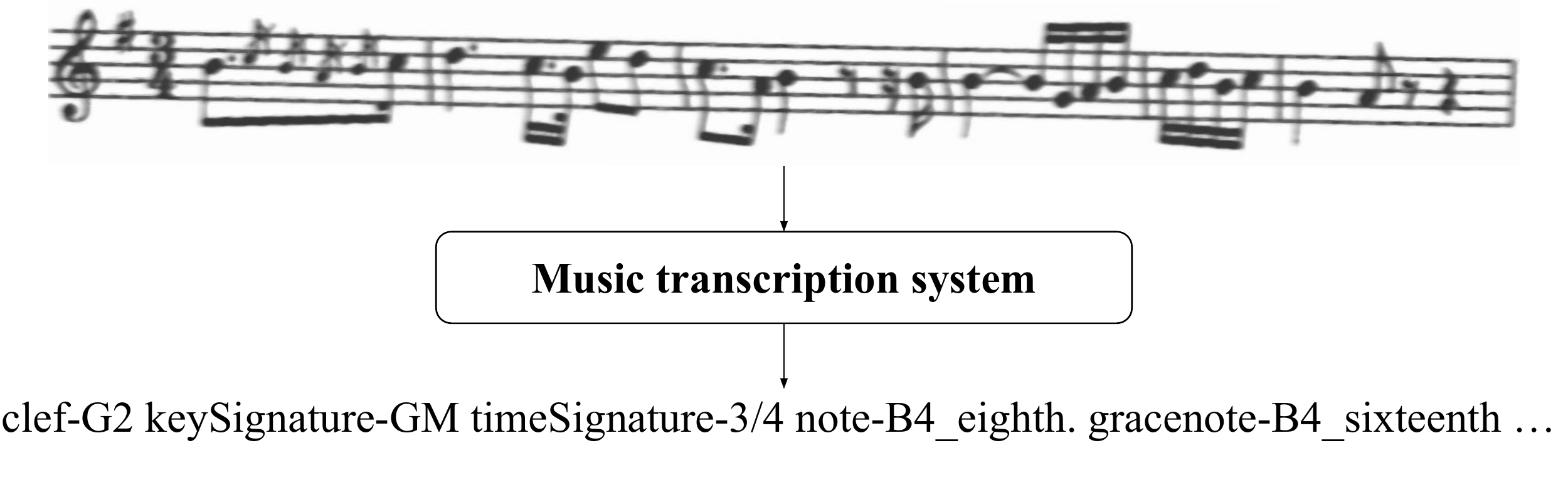}
\label{fig:omr_transcription}}
\hfil
\subfloat[Automatic Music Transcription: music transcription using audio pieces as inputs and the same language as output.]{\includegraphics[width=\columnwidth]{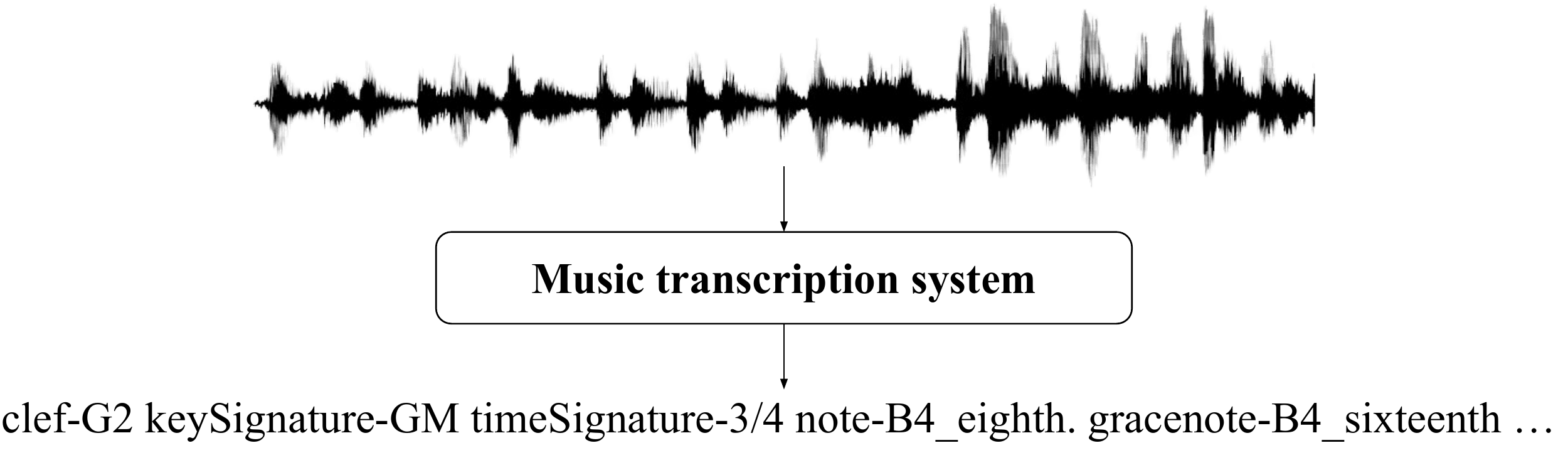}
\label{fig:amt_transcription}}
\caption{End-to-end music transcription framework. OMR techniques deal with images and AMT techniques deal with audio signals; however, both tasks have to provide a result in a symbolic format that represents a piece of music.}
\label{fig:music_transcription}
\end{figure}

However, some recent proposals in MIR literature frame transcription problems in a sequence labeling formulation that approaches the task in a holistic or end-to-end manner~\cite{CalvoZaragoza-Rizo:2018:CameraPrimus, liu2021joint}: the input data---either scores or acoustic pieces---are directly decoded into a sequence of music-notation symbols. This makes it possible to address OMR and AMT tasks with similar recognition models that differ only as regards the input data used to train the system. A graphic illustration of these recognition approaches is provided in Fig.~\ref{fig:music_transcription}.

The common formulation makes it possible to explore possible synergies that may exist between image and audio sources. Some of the possible research avenues yet to be explored by the community are: developing common language models, devising multi-task neural architectures capable of dealing with both tasks independently in a single model, using pre-trained models with one modality and fine-tuning with the other, or multimodal image and audio transcription~\cite{alfaro2022insights}. The last case is the main focus of this work.

Multimodal recognition frameworks, which are defined as those that take multiple representations or modalities of the same piece of data as input, have led to improvements in many fields~\cite{toselli2011multimodal, singh2012improved, pitsikalis2017multimodal}. The different modalities in these schemes are expected to provide complementary information to the system, eventually resulting in an overall improvement in performance. These approaches are generally classified according to the following taxonomy~\cite{dumas2012fusion}: (i) those in which the individual features of the modalities are in some respects directly merged (\textit{feature} or \textit{early-fusion} level), or (ii) those in which the merging process is carried out using the hypotheses obtained by each individual modality (\textit{decision} or \textit{late-fusion} level). Note that the latter family of approaches allows model flexibility, since it requires only the individual hypotheses for the different modalities for the combination.

In the context of music transcription, while this might not be entirely true, both a recording and a score image can be considered as two complementary modalities of the same piece of music. On the one hand, the image provides all the information that was intended to be stored, but is in a graphic domain that must be decoded in order to interpret it as music. On the other, the audio contains information about the musical performance, but certain aspects are difficult to retrieve owing to human interpretation or certain unavoidable ambiguities (such as the clef or the meter). If the original composition were in the form of both an image and a recording, the two could consequently be combined in a synergistic manner so as to exploit the transcription advantages depicted by each modality. A promising task that has, with a few exceptions~\cite{de2021multimodal}, barely been explored in literature, is, therefore, that of developing approaches that are capable of efficiently combining the information conveyed by the image and the audio of the same piece of music for their joint transcription.

In this work, we explore multimodality in image and audio music transcription by combining each individual transcription hypothesis yielded by OMR and AMT models, respectively. More precisely, our contributions are: (i) a thorough revision and first-time application of existing late-fusion recognition approaches to multimodal image and audio music transcription; (ii) comprehensive experimentation with different performance scenarios involving cases in which OMR and AMT depict different base errors, and (iii) the demonstration of a significant improvement to the error rate with respect to existing unimodal approaches.

The remainder of the paper is organized as follows: Section~\ref{sec:background} provides the background to this work, while Section~\ref{sec:unimodal} introduces the proposed approach for unimodal recognition. Section~\ref{sec:multimodal} thoroughly develops the multimodal image and audio transcription framework considered, and Section~\ref{sec:setup} describes the experimental setup. Section~\ref{sec:results} presents and analyses the results, and finally, Section~\ref{sec:conclusions} concludes the work and discusses possible ideas for future research.

\section{Related work}
\label{sec:background}
Traditional attempts to carry out OMR and AMT have been framed within a pipeline that divides the process into a series of independent phases~\cite{Rebelo2012, liu2021audio}. These approaches have been highly conditioned by the nature of their input, thus resulting in incompatible individual (or unimodal) recognition frameworks. However, recent developments have formulated both tasks as \textit{sequence labeling}~\cite{Graves2006}: the input data (either image or audio) is directly decoded into a series of music notation symbols. This is generally accomplished by employing neural end-to-end systems, and specifically Convolutional Recurrent Neural Network (CRNN) architectures owing to their competitive results. In the aforementioned approach, the convolutional stage extracts the most appropriate features for the case in hand, while the recurrent part models the temporal (or spatial) dependencies of these symbols. Given the advantage of being trained using unsegmented sequential data without any input-output alignment, the CRNN scheme introduced is usually considered together with the Connectionist Temporal Classification (CTC) loss function~\cite{Graves2006}. This particular framework currently constitutes one of the state-of-the-art end-to-end approaches for both OMR and AMT tasks~\cite{calvo2020understanding, benetos2018automatic}. This common formulation has naturally triggered an interest in discovering intersections between these research lines, which have previously been addressed separately. Multimodality is, therefore, a suitable avenue of research by which to further study and exploit the synergies between these individual recognition schemes.

Multimodal frameworks seek to leverage the information depicted by each input modality in a synergistic manner~\cite{simonetta2019multimodal}. This combination can be performed at two levels: at the feature or early-fusion level or at the decision or late-fusion level. The former fuses the data ``as is''---the different input modalities undergo a pre-processing phase in order to extract their corresponding features, which are later merged and passed through a single processing algorithm. The latter approach, however, combines the outputs of several ad-hoc algorithms, one for each modality. In this respect, late-fusion multimodality allows more flexibility---particularly when the data sources are significantly different from each other. 

Scientific literature comprises a large number of works that consider recognition tasks within a multimodal framework. A first related example is that by Singh et al.~\cite{singh2012improved}, which proposes the late fusion of individual Text Recognition (TR) and Automatic Speech Recognition (ASR) systems for postal code recognition using a heuristic approach based on the edit distance. In the context of handwritten manuscripts, recent proposals rely on probabilistic frameworks in order to integrate individual estimations using word-graph hypothesis spaces~\cite{granell18_iberspeech} or a compact version of them, namely confusion networks (CN)~\cite{granell2015multimodal}.

Gesture Recognition has also benefited from these multimodal schemes. For instance, Pitsikalis et al.~\cite{pitsikalis2017multimodal} enhance the performance of the model by re-scoring its different hypotheses using information from an ASR system. Other works have considered the use of dynamic programming techniques~\cite{miki2014improvement} or the aforementioned CN paradigm~\cite{kristensson_interspeech2011} to align different hypotheses.

With regard to the MIR field, recent studies exploit different modalities that capture complementary aspects of the same piece of music. In this respect, there are multimodal approaches for different tasks, such as music recommendation, artist identification, or instrument classification, among others~\cite{simonetta2019multimodal}. These techniques are well known in the field of music transcription, and multimodality has been used as a means to break through certain glass-ceiling scenarios reached by single-modality approaches. For instance, research on AMT has contemplated the use of supplementary sources of information, such as onset events, harmonic information, or timbre~\cite{benetos2013automatic}. More recently, the work by de la Fuente et al.~\cite{de2021multimodal} considered that a given score image and its acoustic performance are two different modalities of the same piece to be transcribed. These authors show that transcription results can be enhanced with respect to single-modality systems when their individual performances do not greatly differ.

In this work, we aim to further research the aforementioned avenue of multimodal OMR-AMT transcription. We assume that the score image and the recording of the same piece of music are, in some respects, two complementary sources of information. We follow a late-fusion approach in order to merge these two modalities, as it allows the more adjustable processing of each of them and does not require multimodal training data for the underlying models. More precisely, we study the existing solutions in related areas in order to then adapt them to the focus of the work: image and audio music transcription.

\section{Unimodal neural end-to-end music transcription}
\label{sec:unimodal}
This section formally presents the neural end-to-end recognition framework for the stand-alone OMR and AMT processes. Considering the aforementioned \textit{sequence labeling} formulation, our goal is to retrieve the most likely series of symbols from the input data---either a score image or an audio recording, as appropriate. Some notations will now be introduced in order to properly describe these design principles.

Let $\mathcal{T} = \left\{\left(x_{m},\mathbf{z}_{m}\right) : x_{m}\in\mathcal{X},\;\mathbf{z}_{m}\in\mathcal{Z}\right\}_{m=1}^{|\mathcal{T}|}$ represent a set of data in which sample $x_{m}$ drawn from space $\mathcal{X}$ corresponds to symbol sequence $\mathbf{z}_{m} = \left(z_{m1},z_{m2},\ldots,z_{mN}\right)$ from space $\mathcal{Z}$, considering the underlying function $g : \mathcal{X} \rightarrow \mathcal{Z}$. Note that the latter space is defined as $\mathcal{Z} = \Sigma^{*}$, where $\Sigma$ represents the score-level symbol vocabulary.

We have considered a CRNN scheme with the CTC training algorithm in order to approximate function $g$ owing to its aforementioned competitive performance in the related literature. The CTC method makes it possible to train the CRNN using unsegmented sequential data. In a practical sense, this mechanism requires only the different input signals and their associated sequences of symbols from $\Sigma$ as its expected output, without any specific input-output alignment. It is important to mention that CTC requires an additional ``\textit{blank}'' token within the set of symbols, i.e., $\Sigma' = \Sigma \cup \left\{\textrm{\textit{blank}}\right\}$, to enable the detection of consecutive repeated elements.

During the prediction or decoding phase, CTC assumes that the architecture contains a fully-connected network of $|\Sigma'|$ neurons with a \textit{softmax} activation. Although this inference phase can be performed in several ways~\cite{zenkel2017comparison}, the so-called \textit{greedy} decoding is usually considered. In this respect, assuming that the recurrent layer outputs sequences of length $K$ with $|\Sigma'|$ scores each---namely \textit{posteriorgram}---this process retrieves the most probable symbol per step, as described in Eq.~\ref{eq:ctc-greedy-decoding}.

\begin{equation}
\label{eq:ctc-greedy-decoding}
    \hat{\bm{\pi}} = \arg\max_{\bm{\pi}\in\Sigma'^{K}}{\prod_{k=1}^{K}{y_{\pi_{k}}^{k}}} 
\end{equation}
\noindent where $y_{\pi_{k}}^{k}$ represents the activation probability of symbol $\pi_{k}\in\Sigma'$ and time-step $k$ and $\hat{\bm{\pi}}$ is the retrieved sequence of length $K$.

A $\mathcal{B}\left(\cdot\right)$ mapping function that merges consecutive repeated symbols and removes the \textit{blank} label must eventually be applied to $\hat{\bm{\pi}}$. The predicted sequence is, therefore, obtained as $\hat{\mathbf{z}} = \mathcal{B}\left(\hat{\bm{\pi}}\right)$, where $\left|\hat{\mathbf{z}}\right| \leq K$.

\subsection{Word graphs}
\label{subsec:decoding-wg}
While greedy decoding may be sufficient for unimodal recognition, a policy of this nature may be unsuitable for multimodal transcription based on late-fusion approaches. In this case, it would be appropriate to maintain as much information as possible from the search space---i.e., the set of possible output sequences from the stand-alone recognition models---for the appropriate integration of the different modalities. In this work, we employ lattice-based \textit{Word Graph} (WG) representations owing to their flexibility and successful use in the related literature~\cite{TOSELLI2016497, li2019bi, calvo2019music}. 

\begin{definition*} A WG is a weighted directed acyclic graph, represented as the 5-tuple $w=\left(\Sigma, V, v_{i}, F, E \right)$, where:
\begin{itemize}
    \item $\Sigma$ represents the symbol vocabulary related to the recognition task;
    \item $V$ is a finite set of vertices or nodes, of which $v_{i} \in V$ represents the initial vertex of the graph;
    \item $F \subseteq \left(V - v_{i}\right)$ describes the possible final nodes of the structure; and,
    \item $E \subset \left(V - F\right) \times \left(V - v_{i}\right)$ denotes the finite set of edges, in which each element $e_{A,B} \in E$---edge between vertices $v_{A}$ and $v_{B}$---is labeled with the symbol $l(e_{A,B})\in\Sigma$ and is weighted with a score $s(e_{A,B})\in\mathbb{R}$. Note that this latter value represents the likelihood of symbol $l(e_{A,B})$ appearing between nodes $v_{A}$ and $v_{B}$.
\end{itemize}
\end{definition*}

Using the above as a basis, let $\mathbf{t}=\left(v_1,\dots,v_{|\mathbf{t}|}\right)\mbox{ s.t. } v_{1} = v_{i}\mbox{ and } v_{|\mathbf{t}|}\in F$ denote a complete path within $w$ and $T=\left\{\mathbf{t}_{j}\right\}_{j=1}^{|T|}$ be the set of complete paths in it. Additionally, let $\mathbf{l}_{\mathbf{t}}=\left(l\left(e_{n,n+1}\right)|_{n=1}^{|\mathbf{t}|-1}\right)$ be the sequence of symbols in $\mathbf{t}$ and $s_{\mathbf{t}} =\prod_{n=1}^{|\mathbf{t}|-1} s(e_{n,n+1})$ be its corresponding likelihood.

When considering WG representations, the decoding process selects the most appropriate $\mathbf{t}\in T$ on the basis of certain criteria. This process is usually carried out by considering a \textit{best path} strategy: selecting the sequence that maximizes the likelihood, i.e., $\mathbf{t}^{*} = \arg\max_{\mathbf{t}\in T} s_{\mathbf{t}}$.

In this work, WG representations are obtained by applying a decoding strategy to the posteriorgram output of the aforementioned CRNN model. This decoding is based on Weighted Finite-State Transducers (WFST) and follows the procedure explained in the Kaldi toolkit~\cite{povey2011kaldi}.

\section{Late multimodal fusion}
\label{sec:multimodal}
Late multimodal fusion has principally been considered in the context of recognition systems owing to its reported benefits in terms of performance improvement. In a broad sense, such approaches typically integrate a group of hypotheses into a consensus hypothesis. This work explores the use of decision-level fusion for multimodal neural end-to-end music transcription involving image scores and audio recordings given that the objective in both domains is to retrieve the same symbolic representation. More precisely, we assess the multimodality by combining two respective WG related to an image and an audio recording of the same piece of music. We shall now introduce some notations in order to then formally define this recognition scenario.

Let superscripts $i$ and $a$ identify the image and audio domains, respectively. There are consequently two different representation spaces, $\mathcal{X}^{i}$ and $\mathcal{X}^{a}$, which are related to the image scores and audio signals, respectively, with a single output vocabulary $\Sigma$. Furthermore, let CRNN$^{i}$ and CRNN$^{a}$ denote the two CRNN transcription models introduced in Section~\ref{sec:unimodal} that deal with the source domain of the image and audio data, respectively. 

\begin{figure}[!t]
    \centering
    \includegraphics[width=\columnwidth]{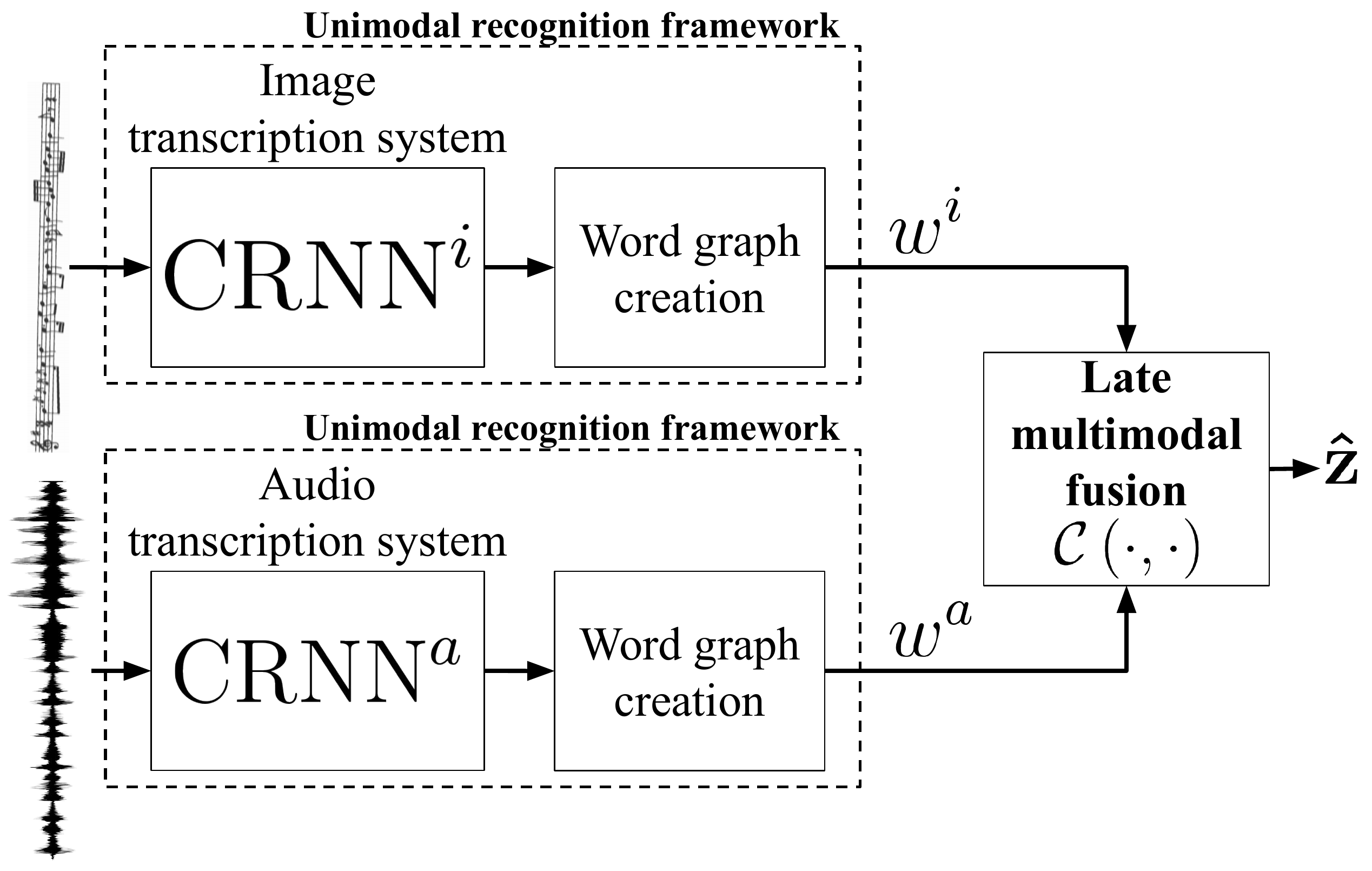}
    \caption{Graphical example of the proposed late multimodal fusion framework: the hypotheses, $w^{i}$ and $w^{a}$, depicted by the CRNN$^{i}$ image and the CRNN$^{a}$ audio transcription models, respectively, are combined according to a certain combination function $\mathcal{C}(\cdot,\cdot)$. This combination results in the symbol sequence eventually predicted: $\hat{\mathbf{z}}$.}
    \label{fig:LF_scheme}
\end{figure}

Given an input pair $\left(x^{i}, x^{a}\right)$ representing a score image and an audio signal of the same piece of music, we obtain their respective $w^{i}$ and $w^{a}$ WG structures by employing the aforementioned $\textrm{CRNN}^{i}$ and $\textrm{CRNN}^{a}$ unimodal recognition methods. Function $\mathcal{C}(\cdot,\cdot)$, which combines two WG and decodes the resulting structure, is then applied to $\left(w^{i}, w^{a}\right)$, and the predicted sequence is obtained as $\hat{\mathbf{z}} = \mathcal{C}\left(w^{i}, w^{a}\right)$. A graphic illustration of this process is provided in Fig.~\ref{fig:LF_scheme}.

In this work, we consider and assess four different $C(\cdot,\cdot)$ combination approaches: (i) one that addresses the fusion carried out by employing the Minimum Bayes Risk criterion~\cite{haihua2011minimum}, (ii) a second case that addresses the task from a lightly-supervised learning perspective~\cite{fainberg2019lattice}, (iii) a third approach that follows a global alignment strategy~\cite{granell2015multimodal}, and (iv) a last procedure based on local alignment~\cite{de2021multimodal}. Their adaptation and first-time application to multimodal image and audio music transcription are described in the following sections.

\subsection{Minimum Bayes Risk approach}
\label{subsec:word-graphs}
The work of Xu et al.~\cite{haihua2011minimum} presents a lattice combination method for speech recognition based on Minimum Bayes Risk (MBR) decoding. More specifically, the authors propose the combination of individual hypotheses from different systems addressing the same task in order to improve the stand-alone recognition rate. In this respect, we can easily adapt this late combination strategy to image and audio music transcription, since CRNN$^{i}$ and CRNN$^{a}$ share a common representation for their expected outputs. 

In WG representations, the objective of MBR decoding is to find the sequence that minimizes the expected risk for a given loss function. In our case, the loss function is the string edit distance, and MBR, therefore, seeks the set median string of the distribution provided by the WG. In late-fusion combination scenarios, MBR performs this search by considering the weighted distributions of the WG being fused. In mathematical terms, for the $m$-th sample in set $\mathcal{T}$, this process is expressed as:
\begin{multline}
\label{eq:mbr-combination}
     \hat{\mathbf{z}}_{m} = \underset{\mathbf{l}_{\mathbf{t}_{m}^{*}}\in\left\{\mathbf{l}_{\mathbf{t}_{m}^{i}}\cup\,\mathbf{l}_{\mathbf{t}_{m}^{a}}\right\}}{\operatorname{argmin}}\left(\alpha \sum_{\mathbf{l}_{\mathbf{t}_{m}^{i}\in T_{m}^{i}}} P(\mathbf{l}_{\mathbf{t}_{m}^{i}} | x_{m}^{i}) \, \textrm{ED}(\mathbf{l}_{\mathbf{t}_{m}^{*}}, \mathbf{l}_{\mathbf{t}_{m}^{i}})\right. + \\
     +\left.\left(1-\alpha\right) \sum_{\mathbf{l}_{\mathbf{t}_{m}^{a}\in T_{m}^{a}}} P(\mathbf{l}_{\mathbf{t}_{m}^{a}} | x_{m}^{a}) \, \textrm{ED}(\mathbf{l}_{\mathbf{t}_{m}^{*}}, \mathbf{l}_{\mathbf{t}_{m}^{a}})\right)
\end{multline}
where $\alpha \in \left(0 , 1\right)$ and $\left(1 - \alpha\right)$ describe the weights given to the image and audio transcription systems, respectively; sets $T_{m}^{i}$ and $T_{m}^{a}$ denote the complete paths for the image and audio recognition methods, respectively;  $\textrm{ED}\left(\cdot,\cdot\right)$ represents the string edit distance~\cite{levenshtein1966binary}; and $P(\mathbf{l}_{\mathbf{t}^{i}_{m}} | x^{i}_{m})$ and $P(\mathbf{l}_{\mathbf{t}^{a}_{m}} | x^{a}_{m})$ denote the respective posterior probabilities of $\mathbf{l}_{\mathbf{t}^{i}_{m}}$ and $\mathbf{l}_{\mathbf{t}^{a}_{m}}$ given the corresponding score image $x^{i}_{m}$ and audio recording $x^{a}_{m}$ samples, which can be approximately computed by following the work of Toselli et al.~\cite{TOSELLI2016497} as:
\begin{equation}
P(\mathbf{l}_{\mathbf{t}_{m}} | x_{m})=\frac{s_{\mathbf{t}_{m}}}{\sum_{\mathbf{t}\in T}s_{\mathbf{t}}}
\end{equation}

In order to study the goodness of this multimodal fusion strategy, we consider different weights in order to assess the balance between image and audio domains in the range of the combination parameter $\alpha\in(0,1)$.

\subsection{Lightly-supervised approach}
\label{subsec:light-word-graphs}
The second late multimodal fusion technique considered in the work was proposed by Fainberg et al.~\cite{fainberg2019lattice}. These authors devised a method with which to correct inaccurate transcriptions by combining them with lattices depicted by a seed recognition model, namely \textit{lightly-supervised} training. In this framework, a transcribed symbol is considered to be correctly estimated if, and only if, it is also present in the lattice of the recognition model.

The method fundamentally depends on a function $f_{\textrm{collapse}}\left(\cdot,\cdot\right)$, which collapses a WG (the compact representation of the hypothesis space of the model) into a sequence of symbols (the transcription) when the aforementioned assumption is met. This results in a new WG with a reduced number of vertices/edges in the matching areas, and remains unaltered otherwise.

We present the adaptation of this decision level multimodal policy to image and audio music transcription in Algorithm~\ref{alg:semi-supervised}. Given an input pair, $\left(x^{(1)}, x^{(2)}\right)$, describing two modality-specific representations of the same source sample, we consider the greedy-decoded hypothesis, $\mathbf{t}^{(1)}$, depicted by CRNN$^{(1)}$, as the transcription to be corrected using the hypothesis, $w^{(2)}$, which is yielded by the domain-counterpart model, CRNN$^{(2)}$. In this work, the two modalities correspond to the image and audio domains. We, therefore, evaluate the technique presented when assuming model correctness in (i)~the image domain, i.e., $w^{(2)} = w^{i}$, and (ii) the audio domain, $w^{(2)} = w^{a}$.

\begin{algorithm}[ht]
\caption{Lightly-supervised late multimodal fusion.}
\label{alg:semi-supervised}
\SetAlgoLined
\SetKwInOut{KwIn}{Input}
\SetKwInOut{KwOut}{Output}
\KwIn{$w^{(1)} \leftarrow$ First WG\\
$w^{(2)} \leftarrow$ Second WG \\
}
\KwOut{$\hat{\mathbf{z}}= \left(\hat{z}_{1},\ldots,\hat{z}_{\left|\hat{\mathbf{z}}\right|}\right) \leftarrow$ Retrieved sequence} 
$\mathbf{l}_{\mathbf{t}^{*(1)}} \leftarrow l(t)\;\forall t \in \arg \max_{\mathbf{t}\in T^{(1)}} s_{\mathbf{t}}$
\Comment{\textit{Best path of $w^{(1)}$}}\\
$w'^{(2)} \leftarrow f_{\textrm{collapse}}\left(\mathbf{l}_{\mathbf{t}^{*(1)}},w^{(2)}\right)$\\
$\hat{\mathbf{z}} \leftarrow l(t)\;\forall t \in \arg \max_{\mathbf{t}\in T'^{(2)}} s_{\mathbf{t}}$
\Comment{\textit{Best path of $w'^{(2)}$}}
\end{algorithm}

\subsection{Global alignment approach}
\label{subsec:confusion-networks}
The work of Granell and Martínez-Hinarejos~\cite{granell2015multimodal} presents a late multimodal fusion approach based on CN with which to combine the outputs of Handwritten Text Recognition (HTR) and ASR systems so as to transcribe handwritten documents. This method can be directly adapted to the image and audio music transcription task by making HTR correspond with OMR and ASR correspond with AMT.

\begin{figure*}[!t]
\centering
\subfloat[Search space as word graph.]{\includegraphics[width=\textwidth]{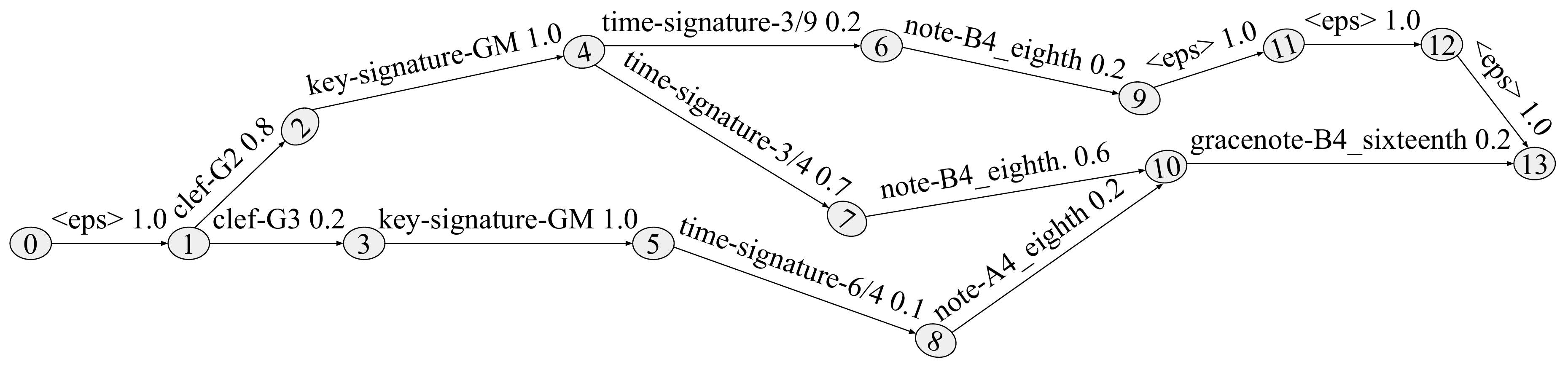}
\label{fig:word-graph}}
\hfil
\subfloat[Search space as confusion network.]{\includegraphics[width=\textwidth]{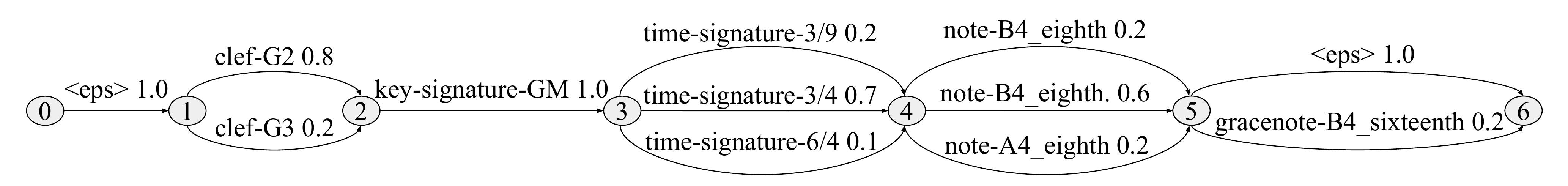}
\label{fig:confusion-network}}
\caption{Search space representation used in the proposed late multimodal fusion framework: Fig.~\ref{fig:word-graph} represents a given $w$ and Fig.~\ref{fig:confusion-network} its corresponding $c$. The symbol sequence ``$<$eps$>$ clef-G2 keySignature-GM timeSignature-3/4 note-B4\_eighth. gracenote-B4\_sixteenth $<$eps$>$'' is the best path ($\mathbf{l}^{*}$) in both lattices.}
\label{fig:lattices}
\end{figure*}

CN is a different topology with which to represent lattice-type spaces. Given $w$, we obtain its corresponding CN, denoted as $c$, by following a clustering process that identifies mutually supporting and competing hypotheses in $w$, and establishing a total order for all of them~\cite{mangu1999finding}. The resulting $c$ is also a weighted directed acyclic graph with three particularities: (i) every path $\mathbf{t} \in T$ from $v_{i}$ to $v_{f}$ passes through all the nodes, i.e., $|\mathbf{t}|=|V|-1 \;\forall \mathbf{t} \in T$, where $v_{f}$ is the final vertex of $c$; (ii) a subnetwork, denoted as $N = \left\{e_{A,B}\right\} \subseteq E$, where $|A-B| = 1$, is the set of all the edges between two consecutive nodes, and (iii) $\sum_{n \in N} s(n) = 1$, i.e., the total score of the edges contained in $N$ adds up to 1. Meeting the first particularity implies dealing with hypotheses of different lengths from $T$. In this respect, one edge (at most), labeled with a special symbol $<$eps$>$, can be inserted into $N$. The inclusion of this symbol results in $c$ adding paths that are not present in $w$. Figure~\ref{fig:lattices} provides a search space formatted as a WG and its equivalent CN.

Given $c^{i}$ and $c^{a}$, the CN-based combination technique seeks to create a new $\hat{c}$ in which the error level is reduced. In a first step, $c^{i}$ and $c^{a}$ are aligned by means of similarity on the basis of a matching error using the Dynamic Time Warping global alignment algorithm~\cite{muller2007dynamic}, where the cost function is the normalized string edit distance between the given the symbols of $N^{i}$ and $N^{a}$. In a second and final step, a new $\hat{c}$ is created by merging subnetworks from the aligned $c^{i}$ and $c^{a}$ by means of Eq.~\ref{eq:cn-combination}:
\begin{equation}
\label{eq:cn-combination}
    s\left(n\right) = s_{\textrm{smoothed}}\left(n^{i}\right)^{\alpha}s_{\textrm{smoothed}}\left(n^{a}\right)^{1-\alpha}
\end{equation}
\noindent where $\alpha \in \left(0 , 1\right)$ and $\left(1 - \alpha\right)$ describe the weights given to the image and audio transcription systems, respectively; and $s_{\textrm{smoothed}}\left(n^{i}\right)$ and $s_{\textrm{smoothed}}\left(n^{a}\right)$ denote the Laplacian smoothed~\cite{zhai2004study} score of edges $n^{i} \in N^{i}$ and $n^{a} \in N^{a}$, respectively.

Please note that if, and only if, the aforementioned normalized string edit distance equals 1---the given symbols of $N^{i}$ and $N^{a}$ are completely different---$N^{i}$ and $N^{a}$ are, respectively, combined with a special subnetwork, $\bar{N}$, that contains only one unit-weighted edge labeled with $<$eps$>$. When the aforementioned condition is not met, $N^{i}$ and $N^{a}$ are combined with each other. The final estimation, $\hat{\mathbf{z}}$, is eventually obtained by retrieving the best path, $\mathbf{t}^{*}$, from the newly-composed $\hat{c}$.

In order to study the balance between the relative reliability of OMR and AMT, we assess the goodness of the late multimodal technique presented when $\alpha \in \left(0, 1\right)$.

\subsection{Local alignment approach}
\label{subsec:smith-waterman}
The last decision level multimodal policy considered in the present work was proposed by de la Fuente et al.~\cite{de2021multimodal}. These authors combined the predictions provided by end-to-end AMT and OMR systems by considering the Smith-Waterman (SW) local alignment algorithm~\cite{SMITH1981195}. 

This method is presented in Algorithm~\ref{alg:sw}. The inputs of the method were originally the hypotheses depicted by each single-modality model decoded by means of Eq.~\ref{eq:ctc-greedy-decoding}. However, in order to maintain consistency with the methodology presented in the current work, we consider the corresponding best paths $\mathbf{t}^{*i}$ and $\mathbf{t}^{*a}$ to be analogous instances of the former greedy-decoded hypotheses. In preliminary experiments carried out for the current work, no differences were found in the recognition results when assuming that this was the case. Given that the $\mathbf{t}^{*i}$ and $\mathbf{t}^{*a}$ may not match in terms of length, the Smith-Waterman (SW) local alignment algorithm~\cite{SMITH1981195} is, therefore, used to align both estimations by searching for the most similar regions between them. The final estimation is eventually obtained from these two aligned sequences, $\bar{\mathbf{t}}^{*i}$ and $\bar{\mathbf{t}}^{*a}$, by following these premises: (i) if both sequences match a symbol, it is included in the resulting estimation; (ii) if they disagree, the symbol with the highest score is included in the estimation, and (iii) if one of the sequences lacks a symbol, that of the other sequence is included in the estimation.

\begin{algorithm}[ht]
\caption{Local alignment late multimodal fusion.}
\label{alg:sw}
\SetAlgoLined
\SetKwInOut{KwIn}{Input}
\SetKwInOut{KwOut}{Output}
\KwIn{$w^{i} \leftarrow$ Image WG\\
$w^{a} \leftarrow$ Audio WG \\
}
\KwOut{$\hat{\mathbf{z}} \leftarrow \left(\hat{z}_{1},\ldots,\hat{z}_{\left|\hat{\mathbf{z}}\right|}\right) $ \Comment{Retrieved sequence}} 
$\mathbf{t}^{*i} \leftarrow \arg \max_{\mathbf{t}\in T^{i}} s_{\mathbf{t}}$ \Comment{\textit{Best path of $w^{i}$}}\\
$\mathbf{t}^{*a} \leftarrow \arg \max_{\mathbf{t}\in T^{a}} s_{\mathbf{t}}$ \Comment{\textit{Best path of $w^{a}$}}\\
$\bar{\mathbf{t}}^{*i}, \bar{\mathbf{t}}^{*a} \leftarrow \textrm{SW}(\mathbf{t}^{*i},\mathbf{t}^{*a})$ \Comment{\textit{SW local alignment}}\\
$K \leftarrow |\bar{\mathbf{t}}^{*i}|$ \Comment{\textit{Sequence length: $|\hat{\mathbf{z}}| = |\bar{\mathbf{t}}^{*i}| = |\bar{\mathbf{t}}^{*a}|$}}\\
\For{$k \in \left[1,\ldots,K\right]$}{
\uIf{$l\left(\bar{t}^{*i}_{k}\right) = \textrm{``''} \vee l\left(\bar{t}^{*i}_{k}\right) = \textrm{``''}$ }{$\hat{z}_{k} \leftarrow l\left(\bar{t}^{*i}_{k}\right) + l\left(\bar{t}^{*a}_{k}\right)$\Comment{\textit{No symbol}}}
\uElseIf{$l\left(\bar{t}^{*i}_{k}\right) = l\left(\bar{t}^{*a}_{k}\right)$}{$\hat{z}_{k} \leftarrow l\left(\bar{t}^{*i}_{k}\right)$\Comment{\textit{Same symbol}}}
\Else{$\hat{z}_{k} \leftarrow l\left(\arg\max_{\bar{t}^{*}_{k}=\left\{\bar{t}^{*i}_{k}, \bar{t}^{*a}_{k}\right\}}s\left(\bar{t}^{*}_{k}\right)\right)$}
}
\end{algorithm}

The same alignment parameters as those employed in the reference work~\cite{de2021multimodal} have been used for the SW algorithm for comparative purposes.

\section{Experimental setup}
\label{sec:setup}
This section presents the definition of the different layers of the neural models, the evaluation protocol used, and the precise evaluation scenarios and corpus considered.

\subsection{Neural network configuration}
\label{subsec:netconfig}
The choice of the CRNN topologies for image and audio music transcription systems is generally conditioned to the particular corpus considered, the amount of accessible data, or the computational resources available, among others. Nevertheless, these configurations usually contain a set of convolutional layers for the feature extraction process, followed by recurrent units for the dependency modeling, with a last dense unit with $|\Sigma'|$ output units. In this work, the actual composition of each layer, depicted in Table~\ref{tab:model-config}, is based on that used in recent works addressing the individual OMR and AMT tasks as a sequence labeling problem~\cite{CalvoZaragoza-Rizo:2018:CameraPrimus, liu2021audio}.

\begin{table*}[!t]
\caption{Layer-wise description of the CRNN models considered. Notation: Conv$(f,w\times h)$ represents a convolution layer of $f$ filters of size $w\times h$ pixels, BatchNorm performs the normalization of the batch, LeakyReLU$(\alpha)$ represents a Leaky Rectified Linear Unit activation with a negative slope of value $\alpha$, MaxPool$(w\times h, a\times b)$ represents the max-pooling operator of dimensions $w\times h$ pixels with $a\times b$ striding factor, BLSTM$(n,d)$ denotes a bidirectional Long Short-Term Memory unit with $n$ neurons and $d$ dropout value parameters, Dense$(n)$ is a fully-connected layer of $n$ neurons, and Softmax($\cdot$) represents the softmax activation. $\Sigma'$ denotes the alphabet considered, including the CTC-blank symbol.}
\label{tab:model-config}
\centering
\resizebox{\textwidth}{!}{
\begin{tabular}{lccccccc}
\toprule[1pt]
    & \textbf{Layer 1} & \textbf{Layer 2} & \textbf{Layer 3} & \textbf{Layer 4} & \textbf{Layer 5} & \textbf{Layer 6} & \textbf{Layer 7}\\ 
    \cmidrule(lr){2-8} 
\textbf{OMR} & \begin{tabular}[c]{@{}l@{}}Conv($64$, $5 \times 5$)\\BatchNorm\\LeakyReLU($0.20$)\\MaxPool($2 \times 2$)\end{tabular} & \begin{tabular}[c]{@{}l@{}}Conv($64$, $5 \times 5$)\\BatchNorm\\LeakyReLU($0.20$)\\MaxPool($1 \times 2$)\end{tabular}    & \begin{tabular}[c]{@{}l@{}}Conv($128$, $3 \times 3$)\\BatchNorm\\LeakyReLU($0.20$)\\MaxPool($1 \times 2$)\end{tabular}   & \begin{tabular}[c]{@{}l@{}}Conv($128$, $3 \times 3$)\\BatchNorm\\LeakyReLU($0.20$)\\MaxPool($1 \times 2$)\end{tabular}     & \begin{tabular}[c]{@{}l@{}}BLSTM($256$)\\Dropout($0.50$)\end{tabular}       & \begin{tabular}[c]{@{}l@{}}BLSTM($256$)\\Dropout($0.50$)\end{tabular}       & \begin{tabular}[c]{@{}l@{}}Dense$(|\Sigma'|)$\\Softmax($\cdot$)\end{tabular}\\ \\
\textbf{AMT} & \begin{tabular}[c]{@{}l@{}}Conv($8$, $2 \times 10$)\\BatchNorm\\LeakyReLU($0.20$)\\MaxPool($2 \times 2$)\end{tabular} & \begin{tabular}[c]{@{}l@{}}Conv($8$, $5 \times 8$)\\BatchNorm\\LeakyReLU($0.20$)\\MaxPool($1 \times 2$)\end{tabular}    & \begin{tabular}[c]{@{}l@{}}BLSTM($256$)\\Dropout($0.50$)\end{tabular}   & \begin{tabular}[c]{@{}l@{}}BLSTM($256$)\\Dropout($0.50$)\end{tabular}     & \begin{tabular}[c]{@{}l@{}}Dense$(|\Sigma'|)$\\Softmax($\cdot$)\end{tabular} & &\\
\bottomrule[1pt]                  
\end{tabular}
}
\end{table*}

All the models in this work were trained using the backpropagation method provided by CTC for $150$ epochs using the ADAM optimizer~\cite{adam} with a fixed learning rate of $0.001$. The batch size was fixed to $16$ for the OMR system, while in the case of the AMT it was set to $4$ because it is more memory-intensive.

\subsection{Performance metrics}
\label{subsec:evaluation}
The performance of the recognition schemes presented is assessed by considering the Symbol Error Rate (SER), as occurred in previous works addressing end-to-end transcription tasks~\cite{CalvoZaragoza-Rizo:2018:CameraPrimus, roman2019holistic}. This figure of merit is computed as the average number of elementary editing operations (insertions, deletions, or substitutions) required in order to match the sequence predicted by the model with that in the ground truth, normalized by the length of the latter. In mathematical terms, this is expressed as:
\begin{equation}
\label{eq:SER}
        \textrm{SER}\;(\%) = \frac{\sum_{m=1}^{|\mathcal{S}|}{\textrm{ED}\left(\mathbf{\hat z}_{m},\;\mathbf{z}_{m}\right)}}{\sum_{m=1}^{|\mathcal{S}|}{|\mathbf{z}_{m}|}}
\end{equation}
where $\mathcal{S}\subset\mathcal{X}\times\mathcal{Z}$ is a set of test data---from either the image or the audio domains---, $\textrm{ED}:\mathcal{Z}\times\mathcal{Z}\rightarrow\mathbb{N}_{0}$ represents the string edit distance, and $\mathbf{\hat z}_{m}$ and $\mathbf{z}_{m}$ denote the estimated and target sequences, respectively.

\subsection{Evaluation scenarios} 
\label{subsec:scenarios}
We aim to provide insights into how the differences among the performances of the stand-alone transcription models affect the outcome of the combined paradigm. The definition of the evaluation scenarios is set according to different ranges of SER. We specifically consider three possible levels:
\begin{itemize}
    \item \textit{High}, which refers to a SER of approximately $30\%$. We consider this performance threshold since ineffective configurations in the related literature report error values similar to this figure. Cases above this level are, therefore, of no interest.
    \item \textit{Low}, which denotes a SER of around $10\%$. As with the previous case, state-of-the-art transcription methods report error rates within this range, and figures lower than this threshold are, therefore, omitted from this study.
    \item \textit{Medium}, which stands for an approximate SER of $20\%$. This intermediate error level has, therefore, been considered for reasons of the completeness of the results.
\end{itemize}

In this respect, we have considered the Camera-based Printed Images of Music Staves (Camera-PrIMuS) database~\cite{CalvoZaragoza-Rizo:2018:CameraPrimus}. This corpus contains 87,678 real music staves of monophonic incipits\footnote{Short sequences of notes, typically the first measures of the piece, used to index and identify a melody or musical work.} extracted from the \textit{Répertoire International des Sources Musicales} (RISM).\footnote{\href{https://rism.info/}{https://rism.info/}} Different representations are provided for each incipit: an image with the rendered score (both plain and with artificial distortions), several encoding formats for the symbol information, and a MIDI file.

Each transcription architecture considers a particular type of data: on the one hand, the OMR model takes the artificially distorted staff image of the incipit as input, and on the other, each MIDI file in the AMT case is synthesized using the FluidSynth software\footnote{\href{https://www.fluidsynth.org/}{https://www.fluidsynth.org/}} and a piano timbre, considering a sampling rate of 22,050 Hz. A time-frequency representation based on the Constant-Q Transform was obtained, with a hop length of 512 samples, 120 bins, and 24 bins per octave, which is eventually embedded as an image that serves as the input. The height of the input considered is scaled to 64 pixels for image data, or to 256 pixels for audio data, maintaining the aspect ratio (signifying that each sample might differ in width) and converted to grayscale, with no further preprocessing. 

Since this corpus was originally devised for OMR tasks, a data cleansing process was carried out in order to adapt it to the multimodal transcription framework presented, resulting in 22,285 incipits.\footnote{This is the case of samples containing long multi-rests, which barely extend the length of the score image but take many frames in the audio signal.} We eventually derive three non-overlapping partitions---train, validation, and test---which correspond to $60\%$, $20\%$, and $20\%$ of the latter amount of data, respectively, following a 5-fold cross-validation scheme. Note that, since the same corpus is considered for both image and audio data, both recognition tasks depict the same label space of $\Sigma^{i} = \Sigma^{a} =$ 1,166 tokens.

When preliminary assessing OMR and AMT models for the original partitions, the resulting error levels are framed within the \textit{High} and \textit{Low} levels, respectively. In order to obtain the other cases, it is consequently necessary to modify the data partitions used in each transcription model. Broadly speaking, since OMR has a base performance framed in the \textit{Low} level, it is possible to obtain the other two cases by simply reducing the train partition; in the case of AMT, however, it is necessary to remove certain incipits from the test partition---those that greatly hinder the performance---as regards achieving the remaining cases. Please note that, since the late-fusion framework addressed in the work requires the same test set for both modalities, the test partition is constrained by AMT.

Considering the design principles presented, the approximate resulting SER rates for the different individual cases are a SER of $27\%$ for the \textit{High} level, a SER of $17\%$ for the \textit{Medium} level, and a SER of $7\%$ for the \textit{Low} level. The combination of those performance standards results in the nine controlled evaluation scenarios used in this work.

\section{Results}
\label{sec:results}
This section presents the results obtained for the proposed experimental scheme. In order to facilitate their analysis, the following notation is introduced for the four late multimodal fusion methods and the nine scenarios considered. On the one hand, \textit{MBR} is used for the MBR-decoding approach (Section~\ref{subsec:word-graphs}); \textit{Lightly}$_{\textrm{\textit{IA}}}$ and \textit{Lightly}$_{\textrm{\textit{AI}}}$ correspond to the lightly-supervised scenario, when assuming model correctness from the audio domain, i.e., the to-be-corrected transcriptions originate from image one and vice versa, respectively (Section~\ref{subsec:light-word-graphs}); \textit{Global} is employed for the global alignment strategy based on CN (Section~\ref{subsec:confusion-networks}), and finally, \textit{Local} corresponds to the SW local alignment procedure (Section~\ref{subsec:smith-waterman}). On the other hand, each scenario is denoted as $\textrm{Scenario N}_{(\textrm{OMR Level Performance - AMT Level Performance})}$, where N denotes the number of the given scenario out of the nine that are possible and \textit{OMR Level Performance} and \textit{AMT Level Performance} refer to the three possible levels of error rate considered---High (H), Medium (M), and Low (L)---for each unimodal model, respectively.\footnote{The code developed in the work is publicly available for reproducible research at: \href{https://github.com/mariaalfaroc/late-fusion-music-transcription.git}{https://github.com/mariaalfaroc/late-fusion-music-transcription.git}}

\begin{figure*}[!t]
    \centering
    \includegraphics[width=\textwidth]{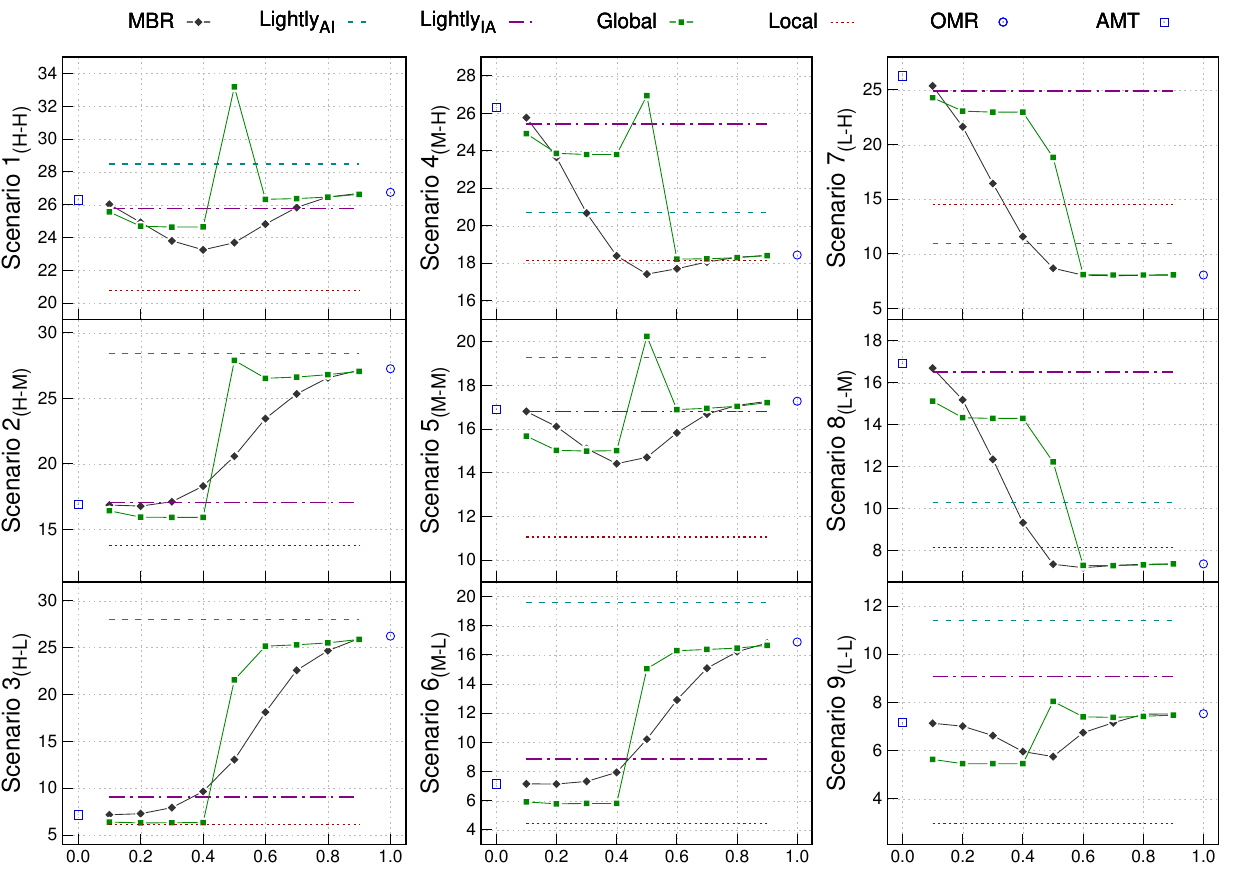}
    \caption{Results obtained by the proposed unimodal and multimodal frameworks in terms of the Symbol Error Rate (SER) for each scenario considered when $\alpha \in \left(0 , 1\right)$. Please note that $\alpha$ is the combination coefficient that allows the weight of each of the unimodal systems involved in the late multimodal fusion to be balanced. $\alpha$ and $\left(1 - \alpha\right)$ specifically describe the weights given to the image and audio transcription systems. $\alpha=1$ and $\alpha=0$ are, therefore, considered to represent the SER figures corresponding to the image and audio domains, respectively.}
    \label{fig:results}
\end{figure*}

The results obtained in terms of the SER metric for the different scenarios considered are presented in Fig.~\ref{fig:results} and Table~\ref{tab:best-results}. Since the experiments have been performed in a cross-validation scheme, the results presented for each of the cases considered correspond to the average values for the test partition in which the validation data achieved the best performance.

\begin{table*}[!t]
\caption{Best results obtained by the proposed unimodal and multimodal frameworks in terms of the Symbol Error Rate (SER) for each scenario considered. For a given scenario, underlining indicates that the corresponding late multimodal fusion approach improves both unimodal baselines and the boldface denotes that it achieves the lowest SER of all the other multimodal methods that also improve the unimodal methods.}
\label{tab:best-results}
\centering
\begin{tabular}{lccccccccc}
\toprule[1pt]
& \multicolumn{9}{c}{\textbf{Scenario}}\\
\cmidrule(lr){2-10}
& \textbf{1$_{\textrm{(H-H)}}$} & \textbf{2$_{\textrm{(H-M)}}$} & \textbf{3$_{\textrm{(H-L)}}$} & \textbf{4$_{\textrm{(M-H)}}$} & \textbf{5$_{\textrm{(M-M)}}$} & \textbf{6$_{\textrm{(M-L)}}$} & \textbf{7$_{\textrm{(L-H)}}$} & \textbf{8$_{\textrm{(L-M)}}$} & \textbf{9$_{\textrm{(L-L)}}$} \\ \cmidrule(lr){1-10}
\textbf{OMR}       & 26.8               & 27.3               & 26.2               & 18.4               & 17.3               & 16.9               & \phantom{0}8.1                & \phantom{0}7.4                & \phantom{0}7.5                \\
\textbf{AMT}       & 26.3               & 16.9               & \phantom{0}7.2                & 26.3               & 16.9               & \phantom{0}7.2                & 26.3               & 16.9               & \phantom{0}7.2                 \\ \cmidrule(lr){1-10}
\textbf{MBR}  & \underline{23.3}               & \underline{16.8}               & \phantom{0}\underline{7.2}                & \underline{\textbf{17.4}}               & \underline{14.4}              & \phantom{0}\underline{7.1}                & \phantom{0}\underline{\textbf{8.0}}       & \phantom{0}\underline{\textbf{7.2}}       & \phantom{0}\underline{5.8}                \\
\textbf{Lightly$_{\textrm{IA}}$}   & \underline{25.8}               & 17.1               & \phantom{0}9.1                & 25.4               & \underline{16.8}               & \phantom{0}8.9                & 24.9               & 16.5               & \phantom{0}9.1                \\
\textbf{Lightly$_{\textrm{AI}}$}  & 28.5               & 28.4               & 28.0                & 20.7               & 19.3               & 19.6                & 10.9               & 10.3               & 11.4                \\
\textbf{Global}  & \underline{24.6}               & \underline{15.9}               & \phantom{0}\underline{6.3}               & \underline{18.2}               & \underline{15.0}                & \phantom{0}\underline{5.8}                & \phantom{0}\underline{8.1}                & \phantom{0}\underline{7.3}               & \phantom{0}\underline{5.5}                 \\
\textbf{Local}  & \underline{\textbf{20.8}}      & \underline{\textbf{13.8}}      & \phantom{0}\underline{\textbf{6.1}}       & \underline{18.2}      & \underline{\textbf{11.1}}      & \phantom{0}\underline{\textbf{4.5}}       & 14.5               & \phantom{0}8.1                & \phantom{0}\underline{\textbf{3.0}}       \\
\bottomrule[1pt]
\end{tabular}
\end{table*}

With regard to the aforementioned results, it is first necessary to state that the image and audio music transcription task can be successfully solved by using a late multimodal fusion approach, as some of the proposed methods employed in the present work achieved lower error rates than their corresponding unimodal baselines for all scenarios. This outcome supports the initial premise that both a recording and an image score can act as two complementary modalities of the same piece of music, since its multimodal combination yields better recognition rates than if performed in an isolated manner. We shall now analyze each method in detail.

As will be noted in both Fig.~\ref{fig:results} and Table~\ref{tab:best-results}, MBR is capable of retrieving improved transcription results while disregarding the actual scenario considered. In relation to the combination coefficient, $\alpha$, there are three distinct situations that are consistent with the differences in the error rate of the models: (i) if both transcription systems are at the same performance level---Scenarios 1, 5, and 9---, there is an almost equal distribution of the weight, slightly shifted towards AMT; (ii) if the image system depicts considerably lower error rates than its audio counterpart---Scenarios 4 and 8---, then the inflection point of the previous weight distribution is now shifted towards the former model, especially when they are at opposite performance levels---Scenario 7---; and finally, (iii) if the audio system is that which works better---Scenarios 2, 3, and 6---, the information provided by that model has a high weight, of at least 80\%, in the combination.

The Lightly method combines an inaccurate transcription---the best-path decoded hypothesis of one modality---with the WG depicted by the seed model---from the other modality. Not surprisingly, the best results provided by this approach are obtained when the corrector is the modality model with the lowest error rate: if the CRNN$^{i}$ image model performs better than its audio counterpart, CRNN$^{a}$, Lightly$_{\textrm{AI}}$ should be used rather than Lightly$_{\textrm{IA}}$, and vice versa. However, both approaches appear to be unsuitable for the image and audio music transcription task as they are generally able only to improve the recognition rate of one of the domain-specific models. On the one hand, Lightly$_{\textrm{IA}}$ is able to enhance transcription results if, and only if, both unimodal baselines depict either a High or a Medium error level. When this condition is not met, it is able only to improve the recognition rate of the worst-performing single-modality transcription system. On the other hand, Lightly$_{\textrm{AI}}$ boosts the audio domain results, at best, when CRNN$^{i}$ is at a lower error rate level than CRNN$^{a}$.

The late multimodal fusion policy proposed by Global behaves in a synergistic manner, as it yields higher performance rates in all scenarios than its corresponding unimodal frameworks. Figure~\ref{fig:results} shows two clear situations related to the balance between the reliability of the image and audio domains, controlled by the combination coefficient $\alpha$. In this respect, when either both unimodal transcription systems are at the same performance level or the audio one has a lower error rate, 70\% of the weight is given to the aforementioned modality. This distribution is reversed when the CRNN$^{i}$ image recognition system outperforms its domain counterpart model, CRNN$^{a}$.

Finally, with regard to Local, the results indicate that the actual improvement achieved by this method depends on the scenario considered. In this respect, this method is not the best multimodal framework when the image system is at a Low error rate level and the audio system is at one of the other two levels (Scenarios 7 and 8). In these cases, it manages only to reduce the error rate of CRNN$^{a}$. However, it is necessary to state that when Local does improve the unimodal recognition rates, it is the best method 90\% of the time, obtaining improvements ranging from 14\% to 58\% with respect to the lowest unimodal error rate. Please note that this upper bound of improvement is obtained when both music transcription systems are at the same performance level. This suggests that this fusion policy has some sort of synergistic behavior in which the resulting sequence takes the most accurate estimations of the OMR and AMT transcription methods.

\subsection{Statistical significance analysis}
\label{subsec:significance}
The analysis shown above can be summarized in the following conclusions:
\begin{itemize}
    \item MBR and Global provide a general improvement with respect to the standard unimodal music transcription framework. When confronting both methods, the former is a better approach than the latter: it yields higher recognition rates in more than half of the scenarios, and is the best performing late multimodal policy in a third of them. However, the differences between them are not remarkable.
    \item Lightly$_{\textrm{IA}}$ improves both unimodal music transcription systems if, and only if, both of them are at the same error level and that level is not Low. However, Lightly$_{\textrm{AI}}$ fails to decrease the error rate of both single-modality models in any scenario.
    \item The Local method improves both domain-specific baselines whenever CRNN$^{i}$ is not at a Low level, because if this is the case, it requires CRNN$^{a}$ to also be at this error level in order to do so. Note that, when possible, it provides the greatest improvement of all the multimodal decision level strategies 90\% of the time.
\end{itemize}

In order to support the relevance of the statements above, we shall now statistically assess the results obtained. This will be done using the non-parametric Wilcoxon signed-rank test~\cite{demvsar2006statistical}. This analysis states that each result obtained for each scenario constitutes a sample of the distributions to be compared. The results obtained are shown in Tables~\ref{tab:wilcoxon-baselines} and~\ref{tab:wilcoxon-methods}.

\begin{table}[!t]
\caption{Statistical significance analysis of the base neural unimodal models of the multimodal fusion schemes considering the Wilcoxon signed-rank test with a significance value of $p < 0.05$ for the Symbol Error Rate metric. Symbols $\checkmark$, $\times$, and $=$ represent that the error of the method in the row is significantly lower than, greater than, or no different to that in the column, respectively.}
\label{tab:wilcoxon-baselines}
\centering
\resizebox{\columnwidth}{!}{
\begin{tabular}{cccccc}
\toprule[1pt]
             & \textbf{MBR} & \textbf{Lightly$_{\textrm{IA}}$} & \textbf{Lightly$_{\textrm{AI}}$} & \textbf{Global} & \textbf{Local} \\ \cmidrule{1-6}
\textbf{OMR} & $\times$               & $=$               & $\checkmark$               & $\times$                & $=$               \\
\textbf{AMT} & $\times$               & $=$                & $=$               & $\times$                & $\times$                \\ \bottomrule[1pt]
\end{tabular}
}
\end{table}

\begin{table}[!t]
\caption{Statistical significance analysis of the late multimodal fusion schemes considering the Wilcoxon signed-rank test with a significance value of $p < 0.05$ for the Symbol Error Rate metric. Symbols $\checkmark$, $\times$, and $=$ represent that the error of the method in the row is significantly lower than, greater than, or no different to that in the column, respectively.}
\label{tab:wilcoxon-methods}
\centering
\resizebox{\columnwidth}{!}{
\begin{tabular}{lccccc}
\toprule[1pt]
         & \textbf{MBR} & \textbf{Lightly$_{\textrm{IA}}$} & \textbf{Lightly$_{\textrm{AI}}$} & \textbf{Global} & \textbf{Local} \\ \cmidrule{1-6}
\textbf{MBR} & -      & $\checkmark$      & $\checkmark$       & $=$      & $=$      \\
\textbf{Lightly$_{\textrm{IA}}$} & $\times$      & -       & $=$      & $\times$      & $\times$      \\
\textbf{Lightly$_{\textrm{AI}}$} & $\times$      & $=$      & -       & $\times$       & $\times$      \\
\textbf{Global} & $=$      & $\checkmark$      & $\checkmark$      & -      & $=$      \\
\textbf{Local} & $=$      & $\checkmark$      & $\checkmark$     & $=$      & -      \\ \bottomrule[1pt]
\end{tabular}
}
\end{table}

The results obtained with a significance value of $p < 0.05$ provide a response to two main questions: (i) is the use of late multimodal fusion frameworks worthwhile?, and (ii) if so, which approach performs best? On the one hand, it is possible to respond to the first by studying Table~\ref{tab:wilcoxon-baselines}, which shows that both MBR and Global significantly outperform both image and audio music transcription systems. It also shows that, while Lightly$_{\textrm{IA}}$ depicts results that are statistically equal to those of both unimodal frameworks, Lightly$_{\textrm{AI}}$ worsens this situation by yielding considerably higher error rates than the OMR model. Finally, Local is able to considerably improve the audio transcription system without hindering the image system. On the other hand, the second question can be answered by studying the results depicted in Table~\ref{tab:wilcoxon-methods}, which shows that there are no meaningful differences among MBR, Global, and Local, but there are with respect to Lightly, in any of its forms, since the three former methods are significantly better than the latter.

Overall, it is now possible to state that the use of late multimodal fusion frameworks is worthwhile for image and audio music transcription, since three out of the four proposed approaches statistically improve at least one of the domain-specific recognition systems without a decline in the performance of the other.
 
\section{Conclusions}
\label{sec:conclusions}
The transcription of music sources into a structured digital format is one of the main challenges in the Music Information Retrieval (MIR) field. This problem is typically divided into two lines of research: Optical Music Recognition (OMR), when considering visual input data such as scores, and Automatic Music Transcription (AMT), when considering acoustic input data such as audio signals. While these fields have historically evolved separately, their recent definition within a sequence labeling formulation results in a common representation for their expected outputs. This enables OMR and AMT tasks to be addressed within a multimodal recognition framework.

In this work, we present the first application of different existing late multimodal fusion approaches to related areas so as to solve the multimodal image and audio music transcription task. We specifically consider four combination functions in order to merge the hypotheses, which are formatted as word graphs, depicted by individual OMR and AMT systems. The results obtained with monophonic music data in a series of performance scenarios---in which the corresponding single-modality models yield different error rates---provided statistical evidence for the benefits of these approaches, since two of the four strategies considered significantly improved the corresponding unimodal standard recognition frameworks.

As future work, we plan to follow different research avenues. For instance, while the approaches described in this paper work at the decision level, it would be interesting to additionally explore early fusion. Experimentation may also be extended to more challenging data, such as handwritten scores, different instrumentation, or polyphonic music.

\section*{Acknowledgments}
This paper is part of the I+D+i PID2020-118447RA-I00 (MultiScore) project, funded by MCIN/AEI/10.13039/501100011033. The first author is supported by grant FPU19/04957 from the Spanish Ministerio de Universidades. The second author is supported by grant APOSTD/2020/256 from ``Programa I+D+i de la Generalitat Valenciana''.

\bibliographystyle{IEEEtran}
\bibliography{bibliography}

\begin{thebibliography}{10}
\providecommand{\url}[1]{#1}
\csname url@samestyle\endcsname
\providecommand{\newblock}{\relax}
\providecommand{\bibinfo}[2]{#2}
\providecommand{\BIBentrySTDinterwordspacing}{\spaceskip=0pt\relax}
\providecommand{\BIBentryALTinterwordstretchfactor}{4}
\providecommand{\BIBentryALTinterwordspacing}{\spaceskip=\fontdimen2\font plus
\BIBentryALTinterwordstretchfactor\fontdimen3\font minus
  \fontdimen4\font\relax}
\providecommand{\BIBforeignlanguage}[2]{{%
\expandafter\ifx\csname l@#1\endcsname\relax
\typeout{** WARNING: IEEEtran.bst: No hyphenation pattern has been}%
\typeout{** loaded for the language `#1'. Using the pattern for}%
\typeout{** the default language instead.}%
\else
\language=\csname l@#1\endcsname
\fi
#2}}
\providecommand{\BIBdecl}{\relax}
\BIBdecl

\bibitem{serra2013roadmap}
X.~Serra, M.~Magas, E.~Benetos, M.~Chudy, S.~Dixon, A.~Flexer,
  E.~G{\'o}mez~Guti{\'e}rrez, F.~Gouyon, P.~Herrera, S.~Jord{\`a}
  \emph{et~al.}, \emph{Roadmap for music information research}.\hskip 1em plus
  0.5em minus 0.4em\relax {The MIReS Consortium}, 2013.

\bibitem{calvo2020understanding}
J.~Calvo-Zaragoza, J.~Haji{\v{c}}~Jr., and A.~Pacha, ``Understanding optical
  music recognition,'' \emph{ACM Computing Surveys (CSUR)}, vol.~53, no.~4, pp.
  1--35, 2020.

\bibitem{benetos2018automatic}
E.~Benetos, S.~Dixon, Z.~Duan, and S.~Ewert, ``{Automatic music transcription:
  An overview},'' \emph{{IEEE Signal Processing Magazine}}, vol.~36, no.~1, pp.
  20--30, 2018.

\bibitem{de2021multimodal}
C.~de~la Fuente, J.~J. Valero-Mas, F.~J. Castellanos, and J.~Calvo-Zaragoza,
  ``Multimodal image and audio music transcription,'' \emph{International
  Journal of Multimedia Information Retrieval}, pp. 1--8, 2021.

\bibitem{CalvoZaragoza-Rizo:2018:CameraPrimus}
J.~Calvo-Zaragoza and D.~Rizo, ``{Camera-PrIMuS: Neural End-to-End Optical
  Music Recognition on Realistic Monophonic Scores},'' in \emph{{Proceedings of
  the 19th International Society for Music Information Retrieval Conference}},
  Paris, France, Sep. 2018, pp. 248--255.

\bibitem{liu2021joint}
L.~Liu, V.~Morfi, and E.~Benetos, ``Joint multi-pitch detection and score
  transcription for polyphonic piano music,'' in \emph{{ICASSP 2021-2021 IEEE
  International Conference on Acoustics, Speech and Signal Processing
  (ICASSP)}}.\hskip 1em plus 0.5em minus 0.4em\relax IEEE, 2021, pp. 281--285.

\bibitem{alfaro2022insights}
M.~Alfaro-Contreras, J.~J. Valero-Mas, J.~M. I{\~n}esta, and J.~Calvo-Zaragoza,
  ``{Insights into transfer learning between image and audio music
  transcription},'' in \emph{{Proceedings of the 19th Sound and Music Computing
  Conference (Accepted)}}.\hskip 1em plus 0.5em minus 0.4em\relax Axea sas/SMC
  Network, 2022.

\bibitem{toselli2011multimodal}
A.~H. Toselli, E.~Vidal, and F.~Casacuberta, \emph{Multimodal interactive
  pattern recognition and applications}.\hskip 1em plus 0.5em minus 0.4em\relax
  Springer Science \& Business Media, 2011.

\bibitem{singh2012improved}
A.~Singh, A.~Sangwan, and J.~H. Hansen, ``Improved parcel sorting by combining
  automatic speech and character recognition,'' in \emph{2012 IEEE
  International conference on emerging signal processing applications}.\hskip
  1em plus 0.5em minus 0.4em\relax IEEE, 2012, pp. 52--55.

\bibitem{pitsikalis2017multimodal}
V.~Pitsikalis, A.~Katsamanis, S.~Theodorakis, and P.~Maragos, ``Multimodal
  gesture recognition via multiple hypotheses rescoring,'' in \emph{Gesture
  recognition}.\hskip 1em plus 0.5em minus 0.4em\relax Springer, 2017, pp.
  467--496.

\bibitem{dumas2012fusion}
B.~Dumas, B.~Signer, and D.~Lalanne, ``{Fusion in multimodal interactive
  systems: an HMM-based algorithm for user-induced adaptation},'' in
  \emph{Proceedings of the 4th ACM SIGCHI symposium on Engineering interactive
  computing systems}, 2012, pp. 15--24.

\bibitem{Rebelo2012}
A.~Rebelo, I.~Fujinaga, F.~Paszkiewicz, A.~Marçal, C.~Guedes, and J.~Cardoso,
  ``Optical music recognition: {S}tate-of-the-art and open issues,''
  \emph{International Journal of Multimedia Information Retrieval}, vol.~1, 10
  2012.

\bibitem{liu2021audio}
L.~Liu and E.~Benetos, ``{From Audio to Music Notation},'' in \emph{{Handbook
  of Artificial Intelligence for Music}}.\hskip 1em plus 0.5em minus
  0.4em\relax Springer, 2021, pp. 693--714.

\bibitem{Graves2006}
A.~Graves, S.~Fern\'{a}ndez, F.~Gomez, and J.~Schmidhuber, ``Connectionist
  {T}emporal {C}lassification: {L}abelling {U}nsegmented {S}equence {D}ata with
  {R}ecurrent {N}eural {N}etworks,'' in \emph{Proceedings of the 23rd
  {I}nternational {C}onference on {M}achine {L}earning}, ser. ICML '06.\hskip
  1em plus 0.5em minus 0.4em\relax New York, NY, USA: ACM, 2006, pp. 369--376.

\bibitem{simonetta2019multimodal}
F.~Simonetta, S.~Ntalampiras, and F.~Avanzini, ``{Multimodal music information
  processing and retrieval: Survey and future challenges},'' in \emph{2019
  International Workshop on Multilayer Music Representation and Processing
  (MMRP)}.\hskip 1em plus 0.5em minus 0.4em\relax IEEE, 2019, pp. 10--18.

\bibitem{granell18_iberspeech}
E.~Granell, C.~D. {Martinez Hinarejos}, and V.~Romero, ``{Improving
  Transcription of Manuscripts with Multimodality and Interaction},'' in
  \emph{Proceedings of IberSPEECH}, 2018, pp. 92--96.

\bibitem{granell2015multimodal}
E.~Granell and C.-D. Martínez-Hinarejos, ``Combining handwriting and speech
  recognition for transcribing historical handwritten documents,'' in
  \emph{{2015 13th International Conference on Document Analysis and
  Recognition (ICDAR)}}, 2015, pp. 126--130.

\bibitem{miki2014improvement}
M.~Miki, N.~Kitaoka, C.~Miyajima, T.~Nishino, and K.~Takeda, ``Improvement of
  multimodal gesture and speech recognition performance using time intervals
  between gestures and accompanying speech,'' \emph{EURASIP Journal on Audio,
  Speech, and Music Processing}, vol. 2014, no.~1, pp. 1--7, 2014.

\bibitem{kristensson_interspeech2011}
P.~O. Kristensson and K.~Vertanen, ``{Asynchronous Multimodal Text Entry using
  Speech and Gesture Keyboards},'' in \emph{Proceedings of the International
  Conference on Spoken Language Processing}, August 2011, pp. 581--584.

\bibitem{benetos2013automatic}
E.~Benetos, S.~Dixon, D.~Giannoulis, H.~Kirchhoff, and A.~Klapuri, ``Automatic
  music transcription: challenges and future directions,'' \emph{Journal of
  Intelligent Information Systems}, vol.~41, no.~3, pp. 407--434, Dec. 2013.

\bibitem{zenkel2017comparison}
T.~Zenkel, R.~Sanabria, F.~Metze, J.~Niehues, M.~Sperber, S.~St{\"u}ker, and
  A.~Waibel, ``{Comparison of Decoding Strategies for CTC Acoustic Models},''
  \emph{Proc. Interspeech 2017}, pp. 513--517, 2017.

\bibitem{TOSELLI2016497}
A.~H. Toselli, E.~Vidal, V.~Romero, and V.~Frinken, ``Hmm word graph based
  keyword spotting in handwritten document images,'' \emph{Information
  Sciences}, vol. 370, pp. 497--518, 2016.

\bibitem{li2019bi}
Q.~Li, P.~Ness, A.~Ragni, and M.~J. Gales, ``{Bi-directional lattice recurrent
  neural networks for confidence estimation},'' in \emph{{ICASSP 2019-2019 IEEE
  International Conference on Acoustics, Speech and Signal Processing
  (ICASSP)}}.\hskip 1em plus 0.5em minus 0.4em\relax IEEE, 2019, pp.
  6755--6759.

\bibitem{calvo2019music}
J.~Calvo-Zaragoza, A.~H. Toselli, E.~Vidal, and J.~A. S{\'a}nchez, ``{Music
  Symbol Sequence Indexing in Medieval Plainchant Manuscripts},'' in
  \emph{{2019 International Conference on Document Analysis and Recognition
  (ICDAR)}}.\hskip 1em plus 0.5em minus 0.4em\relax IEEE, 2019, pp. 882--887.

\bibitem{povey2011kaldi}
D.~Povey, A.~Ghoshal, G.~Boulianne, L.~Burget, O.~Glembek, N.~Goel,
  M.~Hannemann, P.~Motlicek, Y.~Qian, P.~Schwarz \emph{et~al.}, ``The kaldi
  speech recognition toolkit,'' in \emph{IEEE 2011 workshop on automatic speech
  recognition and understanding}, no. CONF.\hskip 1em plus 0.5em minus
  0.4em\relax IEEE Signal Processing Society, 2011.

\bibitem{haihua2011minimum}
H.~Xu, D.~Povey, L.~Mangu, and J.~Zhu, ``{Minimum Bayes Risk Decoding and
  System Combination Based on a Recursion for Edit Distance},'' \emph{Comput.
  Speech Lang.}, vol.~25, no.~4, p. 802–828, oct 2011.

\bibitem{fainberg2019lattice}
J.~Fainberg, O.~Klejch, S.~Renals, and P.~Bell, ``{Lattice-Based
  Lightly-Supervised Acoustic Model Training},'' in \emph{Interspeech 2019,
  20th Annual Conference of the International Speech Communication Association,
  Graz, Austria, 15-19 September 2019}, G.~Kubin and Z.~Kacic, Eds.\hskip 1em
  plus 0.5em minus 0.4em\relax {ISCA}, 2019, pp. 1596--1600.

\bibitem{levenshtein1966binary}
V.~I. Levenshtein, ``Binary codes capable of correcting deletions, insertions,
  and reversals,'' \emph{Soviet physics doklady}, vol.~10, no.~8, pp. 707--710,
  1966.

\bibitem{mangu1999finding}
L.~Mangu, E.~Brill, and A.~Stolcke, ``Finding consensus among words:
  lattice-based word error minimization,'' in \emph{Eurospeech}.\hskip 1em plus
  0.5em minus 0.4em\relax Citeseer, 1999.

\bibitem{muller2007dynamic}
M.~M{\"u}ller, ``Dynamic time warping,'' \emph{Information retrieval for music
  and motion}, pp. 69--84, 2007.

\bibitem{zhai2004study}
C.~Zhai and J.~Lafferty, ``A study of smoothing methods for language models
  applied to information retrieval,'' \emph{ACM Transactions on Information
  Systems (TOIS)}, vol.~22, no.~2, pp. 179--214, 2004.

\bibitem{SMITH1981195}
T.~F. Smith and M.~S. Waterman, ``Identification of common molecular
  subsequences,'' \emph{Journal of Molecular Biology}, vol. 147, no.~1, pp.
  195--197, 1981.

\bibitem{adam}
D.~P. Kingma and J.~Ba, ``{Adam: A Method for Stochastic Optimization},'' in
  \emph{3rd International Conference on Learning Representations, {ICLR} 2015,
  San Diego, CA, USA, May 7-9, 2015, Conference Track Proceedings}, Y.~Bengio
  and Y.~LeCun, Eds., 2015.

\bibitem{roman2019holistic}
M.~A. Rom{\'a}n, A.~Pertusa, and J.~Calvo-Zaragoza, ``A holistic approach to
  polyphonic music transcription with neural networks,'' in \emph{{Proceedings
  of the 20th International Society for Music Information Retrieval
  Conference}}, Delft, The Netherlands, Nov. 2019, pp. 731--737.

\bibitem{demvsar2006statistical}
J.~Dem{\v{s}}ar, ``Statistical comparisons of classifiers over multiple data
  sets,'' \emph{The Journal of Machine Learning Research}, vol.~7, pp. 1--30,
  2006.

\end{thebibliography}

\end{document}